\documentclass[10pt,conference]{IEEEtran}
\usepackage{alltt                                   
	, multirow
	, booktabs
	, listings
	, graphicx
	,float
	,verbatim
	,mathtools
	,url
	,amsmath
}
\usepackage[numbers]{natbib}     
\usepackage{syntax}
\usepackage{algorithm}
\usepackage{enumitem}
\usepackage{threeparttable}
\usepackage{pbox}
\usepackage{algpseudocode}
\usepackage{framed}
\usepackage{hhline}
\usepackage{algpseudocode}
\usepackage{balance}
\usepackage{color, colortbl}
\usepackage{tikz}
\usepackage{verbatim}
\usetikzlibrary{arrows,shapes,backgrounds}

\usepackage{etoolbox}
\usepackage{tikz}
\usetikzlibrary{tikzmark}
\usetikzlibrary{calc}
\usepackage[all]{nowidow}

\tikzstyle{vertex}=[ellipse,fill=black!25,minimum size=20pt, inner sep=0pt]
\tikzstyle{edge} = [draw,thin,-]
\tikzstyle{glabel} = [text width=1cm,text centered,font=\bf]
\pgfdeclarelayer{bg}    
\pgfsetlayers{bg,main}  

\usepackage{expl3}
\ExplSyntaxOn
\newcommand\latinabbrev[1]{
	\peek_meaning:NTF . {
		#1\@}%
	{ \peek_catcode:NTF a {
			#1., \@ }%
		{#1., \@}}}
\ExplSyntaxOff

\algnewcommand\algorithmicswitch{\textbf{switch}}
\algnewcommand\algorithmiccase{\textbf{case}}
\algnewcommand\algorithmicassert{\texttt{assert}}
\algnewcommand\Assert[1]{\State \algorithmicassert(#1)}%
\algdef{SE}[SWITCH]{Switch}{EndSwitch}[1]{\algorithmicswitch\ #1\ \algorithmicdo}{\algorithmicend\ \algorithmicswitch}%
\algdef{SE}[CASE]{Case}{EndCase}[1]{\algorithmiccase\ #1}{\algorithmicend\ \algorithmiccase}%
\algtext*{EndSwitch}%
\algtext*{EndCase}%

\let\footnotesize\scriptsize

\algnewcommand{\LineComment}[1]{\State \(\triangleright\) #1}

\definecolor{LightGray}{rgb}{.9 .9 .9}

\newsavebox{\supbox}
\newcommand{\bsup}{\begin{lrbox}{\supbox}$\tt\scriptstyle}
	\newcommand{\esup}{$\end{lrbox}{}^{\usebox{\supbox}}}
\def\eg{\latinabbrev{e.g}}
\def\ie{\latinabbrev{i.e}}

\definecolor{lightpurple}{rgb}{0.8,0.8,1}
\definecolor{codebg}{RGB}{245,245,245}
\definecolor{commentcolor}{RGB}{11,140,11}
\errorcontextlines\maxdimen

\newcommand{\ALGtikzmarkcolor}{black}
\newcommand{\ALGtikzmarkextraindent}{4pt}
\newcommand{\ALGtikzmarkverticaloffsetstart}{-.5ex}
\newcommand{\ALGtikzmarkverticaloffsetend}{-.5ex}
\makeatletter
\newcounter{ALG@tikzmark@tempcnta}

\newcommand\ALG@tikzmark@start{%
	\global\let\ALG@tikzmark@last\ALG@tikzmark@starttext%
	\expandafter\edef\csname ALG@tikzmark@\theALG@nested\endcsname{\theALG@tikzmark@tempcnta}%
	\tikzmark{ALG@tikzmark@start@\csname ALG@tikzmark@\theALG@nested\endcsname}%
	\addtocounter{ALG@tikzmark@tempcnta}{1}%
}

\def\ALG@tikzmark@starttext{start}
\newcommand\ALG@tikzmark@end{%
	\ifx\ALG@tikzmark@last\ALG@tikzmark@starttext
	\else
	\tikzmark{ALG@tikzmark@end@\csname ALG@tikzmark@\theALG@nested\endcsname}%
	\tikz[overlay,remember picture] \draw[\ALGtikzmarkcolor] let \p{S}=($(pic cs:ALG@tikzmark@start@\csname ALG@tikzmark@\theALG@nested\endcsname)+(\ALGtikzmarkextraindent,\ALGtikzmarkverticaloffsetstart)$), \p{E}=($(pic cs:ALG@tikzmark@end@\csname ALG@tikzmark@\theALG@nested\endcsname)+(\ALGtikzmarkextraindent,\ALGtikzmarkverticaloffsetend)$) in (\x{S},\y{S})--(\x{S},\y{E});%
	\fi
	\gdef\ALG@tikzmark@last{end}%
}

\apptocmd{\ALG@beginblock}{\ALG@tikzmark@start}{}{\errmessage{failed to patch}}
\pretocmd{\ALG@endblock}{\ALG@tikzmark@end}{}{\errmessage{failed to patch}}
\makeatother

\begin{document}	
	\title{Effective Reformulation of Query for Code Search using  Crowdsourced Knowledge and Extra-Large Data Analytics\vspace{-.4cm}}
	
	\author{\IEEEauthorblockN{Mohammad Masudur Rahman  ~~~ Chanchal K. Roy}
		\IEEEauthorblockA{Department of Computer Science, University of Saskatchewan, Canada\\
			\{masud.rahman, chanchal.roy\}@usask.ca}
	}
	
	\maketitle
	
	\begin{abstract}
		Software developers frequently issue generic natural language queries for code search while using code search engines (\eg\ GitHub native search, Krugle).  
		Such queries often do not lead to any relevant results due to vocabulary mismatch problems.
		In this paper, we propose a novel technique that 
		automatically identifies relevant and specific API classes from Stack Overflow Q \& A site for a programming task written as a natural language query, and then reformulates the query for improved code search. 
		We first collect candidate API classes from Stack Overflow  using pseudo-relevance feedback and two term weighting algorithms, and then rank the candidates using Borda count and semantic proximity between query keywords and the API classes. The semantic proximity has been determined by an analysis of 1.3 million questions and answers of Stack Overflow. 
		Experiments using 310 code search queries report that our technique suggests relevant API classes with 48\% precision and 58\% recall which are 32\% and 48\% higher respectively than those of the state-of-the-art. 
		Comparisons with two state-of-the-art studies and three popular search engines (\eg\ Google, Stack Overflow, and GitHub native search) report that our reformulated queries (1) outperform the queries of the state-of-the-art, and (2) significantly improve the code search results provided by these contemporary search engines. 
		
	\end{abstract}




\IEEEpeerreviewmaketitle

\section{Introduction}\label{sec:introduction}
Software developers spend about 19\% of their development time in searching for relevant code snippets (\eg\ API usage examples) on the web \cite{twostudy}. Although open source software repositories (\eg\ GitHub, SourceForge) are a great source of such code snippets, retrieving them is a major challenge \cite{sourcerer}. 
Developers often use code search engines (\eg\ Krugle, GitHub native search) to collect code snippets from such repositories using generic natural language queries \cite{koderlog}. 
Unfortunately, such queries hardly lead to any relevant results (\ie\ only 12\% \cite{koderlog}) due to vocabulary mismatch issues \cite{vocaprob,sitir}. 
Hence, the developers frequently reformulate their queries by removing irrelevant keywords and by adding more appropriate keywords. 
Studies \cite{kevic,stolee2015,koderlog} have shown that 33\%--73\% of all the queries are 
incrementally reformulated by the developers.
These manual reformulations involve numerous trials and errors, and often cost significant development time and efforts \cite{kevic}. 
One way to help the developers overcome this challenge is to automatically reformulate their
generic queries (which are often poorly designed \cite{kevic,sitir}) using appropriate query keywords such as relevant API classes. 
Our work in the paper addresses this particular research problem -- \emph{query reformulation} targeting \emph{code search}.

Several existing studies offer automatic query reformulation supports for code search using either actual or pseudo relevance feedback on the query \cite{tsc2016,active} and by mining crowd generated knowledge stored in Stack Overflow programming Q \& A site \cite{tsc2016,li2016,rack}. \citet{tsc2016} collect pseudo-relevance feedback (PRF) on a given query by employing Stack Overflow as a feedback provider, and then suggest query expansion by analysing the feedback documents, \ie\ relevant   programming questions and answers. However, they treat the Q \& A threads as regular texts, and suggest natural language (\ie\ software-specific) terms as query expansion. 
Existing evidence suggests that queries containing only natural language terms perform poorly in code search \cite{koderlog}.
\citet{saner2016masud} mine co-occurrences between query keywords (found in the question titles) and API classes (found in the answers) of Stack Overflow, apply two heuristics, and then suggest a set of relevant API classes for a given query. 
Unfortunately, their heuristics are likely to return highly \emph{generic} and \emph{frequent} API classes (\eg\ \texttt{String, ArrayList, List}) due to their sole reliance on the co-occurrences. They even might return false positives given that all Q \& A threads from the corpus were used for each query rather than the relevant ones.

In this paper, we propose a novel technique--NLP2API--that automatically identifies relevant API classes for a programming task written as a natural language query, and then reformulates the query using these API classes for improved code search.
We first (1) collect candidate API classes for a query from relevant 
questions and answers of Stack Overflow (\ie\ \emph{crowdsourced knowledge}) (Section \ref{sec:candidate}), and then (2) identify appropriate classes from the candidates using Borda count (Section \ref{sec:borda}) and query-API semantic proximity (\ie\ \emph{extra-large data analytics}) (Section \ref{sec:w2vec}).
In particular, we determine semantics of either a keyword or an API class based on their positions within a high dimensional semantic space developed by \textit{fastText} \cite{fasttext} using 1.3 million questions and answers of Stack Overflow. 
Then we estimate the relevance of the candidate API class to the search query using their semantic proximity measure.
Earlier approaches only perform either local context analysis \cite{tsc2016,active} or global context analysis \cite{saner2016masud,pwordnet,thesaurus}. On the contrary, our technique analyses both local (\eg\ PageRank \cite{pagerank}) and global (\eg\ semantic proximity) contexts of the query keywords for  relevant API class identification and query reformulation. Thus, NLP2API has a higher potential for query reformulation. Besides, opportunistic blending of pseudo-relevance feedback \cite{prf,qsurvey}, term weighting methods \cite{pagerank,tfidf}, Borda count \cite{borda1} and \emph{extra-large data analytics} \cite{fasttext} also makes our work in this paper \emph{novel}. 

%
%

\begin{lstlisting}[label=lst:example, language=java,  escapechar=@, aboveskip=1em, float=t, belowskip=-1em, frame=bt, caption={An example code snippet for the programming task-- ``Convert image to grayscale without losing transparency", (taken from \cite{examplecode})}]
 @\textbf{BufferedImage}@ master = ImageIO.read(new URL(
 "http://www.java2s.com/style/download.png"));
 BufferedImage gray = new BufferedImage(master.getWidth(),
 master.getHeight(), BufferedImage.TYPE_INT_ARGB);

 @\textbf{ColorConvertOp}@ op = new ColorConvertOp(
 @\textbf{ColorSpace}@.getInstance(ColorSpace.CS_GRAY), null);
 op.filter(master, gray);

 ImageIO.write(master,"png",new File("path/to/master"));
 ImageIO.write(gray,"png", new File("path/to/gray/image"));
\end{lstlisting}

\begin{table}[!t]
	\centering
	\caption{Reformulations of an NL Query for Improved Code Search}
	\label{table:exampleref}
	\vspace{-.2cm}
	\resizebox{3.4in}{!}{%
		\begin{threeparttable}
			\begin{tabular}{l|p{2.4in}|c}
				\hline
				\textbf{Technique} & \textbf{Reformulated Query} & \textbf{QE}\\
				\hline
				\hline
				Baseline & Convert image to grayscale without losing transparency & 115 \\
				\hline
				QECK \cite{tsc2016} & \{Convert image grayscale losing	transparency\} + \{hsb pixelsByte png iArray img correctly HSB mountainMap enhancedImagePixels file\} & \textbf{11}  \\
				\hline
				Google & Convert image to grayscale without losing transparency & \textbf{02}\\ 
				\hline
				& \{Convert image grayscale losing	transparency\} + \{\texttt{\textbf{BufferedImage} Grayscale ImageEdit}\\ 
				\textbf{Proposed} & \texttt{\textbf{ColorConvertOp} File Transparency \textbf{ColorSpace} BufferedImageOp Graphics ImageEffects}\} & \textbf{02}\\
				\hline
			\end{tabular}
		\centering
		\textbf{QE} = Rank of the first correct result returned by the query
		\end{threeparttable}
	}
	\vspace{-.6cm}
\end{table}

Table 1 and Listing \ref{lst:example} present a use-case scenario of our technique where a developer is looking for a working code snippet that can 
convert a colour image to grayscale without losing transparency. 
First, the developer issues a generic query--\emph{``Convert image to grayscale without losing transparency"}. Then she submits it to \emph{Lucene}, a search engine that is widely used both by contemporary code search solutions such as GitHub native search \cite{gh-lucene} or ElasticSearch and by academic studies \cite{refoqus,saner2017masud,trconfig}. Unfortunately, the generic natural language query does not perform well due to vocabulary mismatch between its keywords and the source code, and returns the relevant code snippet (\eg\ Listing \ref{lst:example}) at the \textbf{115$^{th}$} position. On the contrary, (1) our proposed technique \emph{complements} this query with not only \emph{relevant}, but also highly \emph{specific} API classes (\eg\ \texttt{\textbf{BufferedImage}}, \texttt{\textbf{ColorConvertOp}}, \texttt{\textbf{ColorSpace}}), and (2) our improved query returns the target code snippet at the \textbf{\emph{second}} position of the ranked list which is a major rank improvement over the baseline. The most recent and closely related technique--QECK \cite{tsc2016}
returns the same code snippet at the 11$^{th}$ position which is not ideal. 
Google, the most popular web search engine, returns a \emph{similar} code at the second position as well. However, in the case of web search, relevant code snippets are \emph{sporadic} and often buried into a large bulk of unstructured, noisy and redundant natural language texts across multiple web pages which might overwhelm the developer with information overload \cite{cexchange}.


Experiments using 310 code search queries randomly collected from four Java tutorial sites--\emph{KodeJava, Java2s, CodeJava} and \emph{JavaDB}--report that our technique can suggest relevant API classes with 82\% Top-10 Accuracy, 48\% precision, 58\% recall and a reciprocal rank of 0.55 which are 6\%, 32\%, 48\% and 41\% higher respectively than those of the state-of-the-art \cite{saner2016masud}. 
Comparisons with three state-of-the-art studies and three popular code (or web) search engines -- \emph{Google}, \emph{Stack Overflow native search} and \emph{GitHub native search} -- reported that our technique (1) can outperform the existing studies \cite{tsc2016,cocabu,saner2016masud} in query effectiveness and (2) can improve upon the precision of these search engines by 17\%, 34\% and 33\% respectively using our reformulated queries.
Thus, this paper makes the following contributions:

\begin{itemize}[noitemsep,topsep=0pt]
\item A novel technique that reformulates generic natural language queries for improved code search using large data analytics and crowd knowledge stored in Stack Overflow.
\item  Comprehensive evaluation of the proposed technique using 310 queries and validation against the state-of-the-art techniques and widely used code search engines.
\item A replication package that includes our working prototype and the detail experimental dataset.
\end{itemize}

       
 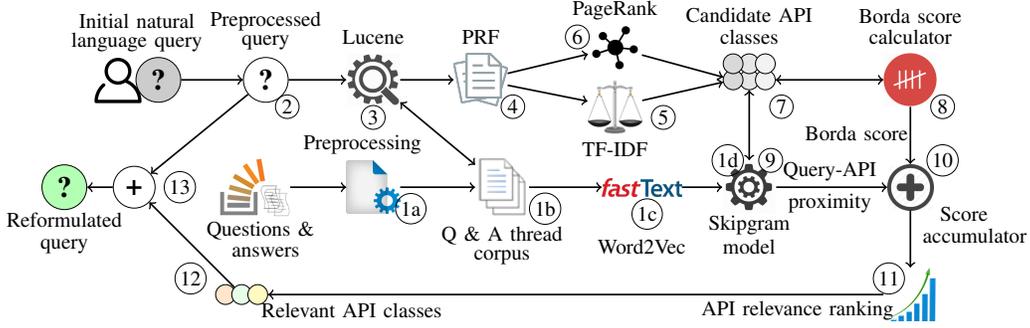
\begin{figure*}
 	\centering
 	\resizebox{5.45in}{!}{%
 		\begin{tikzpicture}[scale=.9, auto,swap]
 		
 		\begin{pgfonlayer}{bg}
 		\node[inner sep=0pt] (posts) at (-2.2,-2)
 		{\includegraphics[scale=.25]{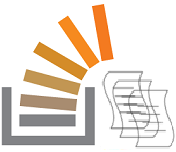}};
 		\node at (-2.1,-2.8) (b) {Questions \& };
 		\node at (-2,-3.2) (b) {answers};

 		\node[inner sep=0pt] (docs) at (2,0)
 		{\includegraphics[width=.35in]{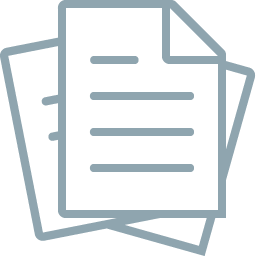}};
 		\node at (2,.8) (b) {PRF};
 		
 		 \node[inner sep=0pt] (wn) at (4.5,0.6)
 		{\includegraphics[width=.35in]{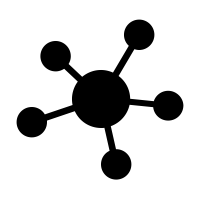}};
 		\node at (4.5,1.3) (b) {PageRank};

 		\node[inner sep=0pt] (ce) at (0,-2)
 		{\includegraphics[width=.35in]{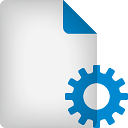}};
 		\node at (-.2,-1.2) (b) {Preprocessing};
 		\draw[->,thick] (posts) -- (ce);
 		
 		\node[inner sep=0pt] (ppce) at (2.4,-2)
 		{\includegraphics[width=.35in]{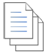}};
 		\draw[->,thick] (ce) -- (ppce);
 		\node at (2.4,-2.9) (b) {Q \& A thread};
 		\node at (2.4,-3.3) (b) {corpus};

 		\node[inner sep=0pt] (lucene) at (0,0)
 		{\includegraphics[width=.35in]{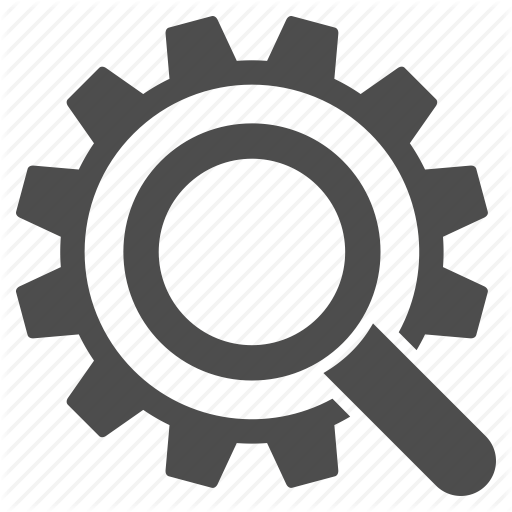}};
 		\node at (0,.8) (b) {Lucene};
 		\draw[->,thick] (lucene) --(docs);
 		\draw[<->,thick] (ppce) --(lucene);
 		
 	    \node at (-2,0) [circle,draw] (pp) {\Large \textbf{?}};
 	    \draw[->,thick] (pp) -- (lucene);
 	    \node at (-2,1.1) (b) {Preprocessed};
 	    \node at (-2,.7) (b) {query};
 	    
 	    \node at (-4,0) [circle,draw, fill=gray!40] (initq) {\Large \textbf{?}};
 	    \draw[->,thick] (initq) -- (pp);
 		\node[inner sep=0pt] (user) at (-4.75,0)
 		{\includegraphics[width=.35in]{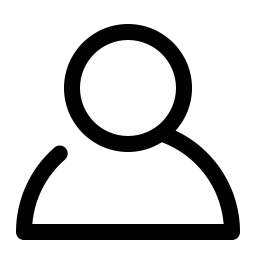}};
 		\node at (-4.4,1.1) (b) {Initial natural};
 		\node at (-4.4,.7) (b) {language query};
 		

	 	 \node[inner sep=0pt] (tfidf) at (4.5,-.5)
	 	{\includegraphics[scale=.20]{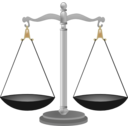}};
	 	\node at (4.5,-1.3) (b) {TF-IDF};

 		\draw[->,thick] (docs) -- (wn);
 		\draw[->,thick] (docs) -- (tfidf);

 		\node[inner sep=0pt] (fasttext) at (5,-2)
 		{\includegraphics[scale=.5]{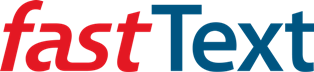}};
 		\node at (5,-3.1) (b) {Word2Vec};

 		\node[inner sep=0pt] (model) at (7,-2)
 		{\includegraphics[width=.35in]{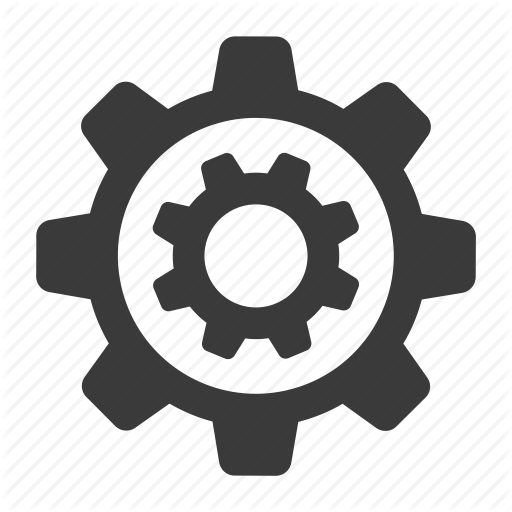}};
 		\node at (7,-2.7) (b) {Skipgram};
 		\node at (7,-3.1) (b) {model};
 		\draw[->,thick] (fasttext) -- (model);
 		\draw[->,thick] (ppce) -- (fasttext);

 		\node at (7,1.2) (b) {Candidate API};
 		\node at (7,.8) (b) {classes};
 		\node at (7,0)(c4) [circle,draw, fill=gray!20]{};
 		\node at (7.3,0)(c2) [circle,draw, fill=gray!10]{};
 		\node at (6.7,0)(c3) [circle,draw, fill=gray!30]{};
 		\node at (7,.3)(c1) [circle,draw, fill=gray!20]{};
 		\node at (7.3,.3) [circle,draw, fill=gray!10]{};
 		\node at (6.7,.3)(cb) [circle,draw, fill=gray!30]{};

 		\draw[->,thick] (tfidf) -- (c3);
 		 		\draw[->,thick] (wn) -- (c3);
 		\draw[<->,thick] (model) -- (c4);
 		
 		 \node[inner sep=0pt] (borda) at (10,0)
 		{\includegraphics[width=.35in]{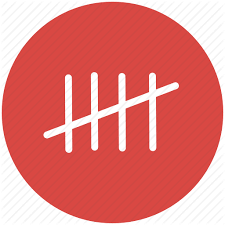}};
 		\node at (10,1.2) (b) {Borda score};
 		\node at (10,.8) (b) {calculator};
 		\draw[<->,thick] (borda) -- (c2);

 		 \node[inner sep=0pt] (ranking) at (10,-2)
 		{\includegraphics[width=.30in]{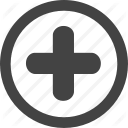}};
 		\draw[->,thick] (borda) -- (ranking);
 		\draw[->,thick] (model) -- (ranking);
 		\node at (8.5,-1.7) (b) {Query-API};
 		\node at (8.5,-2.3) (b) {proximity};
 		\node at (9,-1) (b) {Borda  score};
 		\node at (11,-2.5) (b) {Score};
 		\node at (11.2, -2.9)(b){accumulator};

 		\node[inner sep=0pt] (ranked) at (10,-4)
 		{\includegraphics[width=.35in]{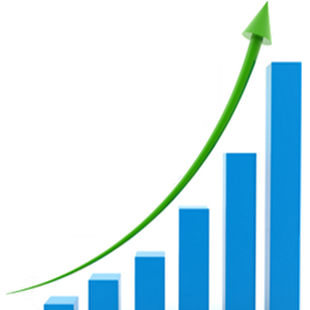}};
 		\draw[->,thick] (ranking) -- (ranked);
 		\node at (7.9,-4.3) (b) {API relevance ranking};

 		\node at (-2.75,-4)(c6) [circle,draw, fill=orange!20]{};
 		\node at (-2.45,-4)(c7) [circle,draw, fill=green!10]{};
 		\node at (-2.15,-4)(c5) [circle,draw, fill=yellow!30]{};
 		\draw[->,thick] (ranked) -- (c5);
 		\node at (-.4,-4.3) (b) {Relevant API classes};

 		\node at (-4.45,-2)(add) [circle,draw]{\Large\textbf{+}};
 		\draw[->,thick] (c7) -- (add);
 		\draw[->,thick] (pp) -- (add);
 			
 		\node at (-5.75,-2)(ref) [circle,draw,fill=green!30]{\Large \textbf{?}};
 		\draw[->,thick] (add) -- (ref);
 		\node at (-5.75,-2.6) (b) {Reformulated};
 		\node at (-5.75,-3.1) (b) {query};

 		\node[inner sep=1pt] at (.7,-2.4) [circle,draw] (b) {1a};
 		\node[inner sep=1pt] at (3.2,-2.4) [circle,draw] (b) {1b};
 		\node[inner sep=1pt] at (5.1,-2.5) [circle,draw] (b) {1c};
 		\node[inner sep=1pt] at (6.6,-1.5) [circle,draw] (b) {1d};
 		\node[inner sep=1pt] at (-1.6,-.5) [circle,draw] (b) {2};
 		\node[inner sep=1pt] at (0,-.7) [circle,draw] (b) {3};
 		\node[inner sep=1pt] at (2.6,-.5) [circle,draw] (b) {4};
 		\node[inner sep=1pt] at (3.8,.8) [circle,draw] (b) {6};
 		\node[inner sep=1pt] at (5.4,-.7) [circle,draw] (b) {5};
 		\node[inner sep=1pt] at (7.6,-.5) [circle,draw] (b) {7};
 		\node[inner sep=1pt] at (10.6,-.5) [circle,draw] (b) {8};
 		\node[inner sep=1pt] at (7.4,-1.5) [circle,draw] (b) {9};
 		\node[inner sep=1pt] at (10.6,-1.5) [circle,draw] (b) {10};
 		\node[inner sep=1pt] at (9.6,-3.7) [circle,draw] (b) {11};
 		\node[inner sep=1pt] at (-3.4,-3.7) [circle,draw] (b) {12};
 		\node[inner sep=1pt] at (-3.6,-2) [circle,draw] (b) {13};

 		 		
 		\end{pgfonlayer}
 		\end{tikzpicture}
 	}
 	\vspace{-.3cm}
 	\caption{Schematic diagram of the proposed query reformulation technique--NLP2API}
 	\label{fig:sysdiag}
 	\vspace{-.6cm}
 \end{figure*}

\section{Background}
\label{sec:bg}
\textbf{Pseudo-Relevance Feedback (PRF):} 
\citet{gayg} first use relevance feedback for query reformulation in the context of Software Engineering, more specifically in concept location. They first collect a developer's feedback on a given query where the developer marks each result returned by the query as either \emph{relevant} or \emph{irrelevant}. Then they analyse the marked results, extract appropriate keywords using Rocchio's method \cite{rocchio}, and reformulate the given query. Although the interactive feedbacks from the developer are effective, collecting them is time-consuming and sometimes infeasible \cite{manning-ir}. Hence, researchers adopted a less effective but more feasible approach namely \emph{pseudo-relevance feedback} \cite{refoqus,tsc2016,ase2017amasud}. In this approach, only the Top-K results (returned by the given query) are assumed as relevant and thus, are \emph{automatically} analysed for the reformulation task. Studies have shown that pseudo-relevance feedback based reformulations could improve the initial queries significantly (\ie\ $\approx$ 20\%) \cite{qsurvey}.   
In our work, we adopt pseudo-relevance feedback as a step for candidate API class selection from Stack Overflow Q \& A threads.

\textbf{Word Embeddings (WE):} 
Traditional code search engines often suffer from vocabulary mismatch issues (\eg\ polysemy, synonymy) \cite{sitir,infer,qeffect}. One crucial step to overcome this challenge is to determine semantics of a word correctly.
There have been several attempts \cite{w2vec,wordsim,wembedding,ase2016masud} to define the semantics of a word by using its contexts captured from a large corpus (\eg\ Stack Overflow). \citet{w2vec} propose \emph{word2vec}, a feed-forward neural network based text mining tool that mines a corpus, and represents each word as a point within a high-dimensional semantic space. 
Thus, the semantics of each word is represented as a vector, and similar words occur close to each other on the semantic space. 
Such vector is also called \emph{word embeddings} \cite{wembedding,w2vec}. \emph{word2vec} is trained using two predictive models-- continuous bag of words (CBOW) and skip-gram. CBOW model predicts a word given its contextual words whereas skip-gram attempts to predict the context of a given word probabilistically. 
We use a faster version of \emph{word2vec} called \emph{fastText} \cite{fasttext} with its default parameters for query reformulation in this paper. 

\section{NLP2API: Proposed Technique for Query Reformulation for Improved Code Search}    
Fig. \ref{fig:sysdiag} shows the schematic diagram of our proposed technique for the reformulation of a generic query targeting code search. Furthermore,  
Algorithm \ref{algo} shows the pseudo-code of our technique. We make use of pseudo-relevance feedback (PRF), crowd generated knowledge stored at Stack Overflow, two term weighting algorithms, and extra-large data analytics for our query reformulation as follows:  


\subsection{Development of Candidate API Lists}\label{sec:candidate}
We collect candidate API classes from Stack Overflow Q \& A site to reformulate a generic query (\ie\ Fig. \ref{fig:sysdiag}, Steps 1a, 1b, 2--7). 
Stack Overflow is a large body of crowd knowledge with 14 million questions and 22 million answers across multiple programming languages and domains \cite{nlp2code}. Hence, it might contain at least a few questions and answers related to any programming task at hand. 
Earlier studies from the literature \cite{tsc2016,li2016,nlp2code} also strongly support this conjecture. Given that relevant program elements are a better choice than generic natural language terms for code search \cite{koderlog}, we collect API classes as candidates for query reformulation by mining the programming Q \& A threads of Stack Overflow.

\textbf{Corpus Preparation:} We collect a total of 656,538 Q \& A threads related to \texttt{Java} (\ie\ using \texttt{<java>} tag) from Stack Overflow for corpus preparation (Fig. \ref{fig:sysdiag}, Steps 1a, 1b, Algorithm \ref{algo}, Line 3). We use the public data dump \cite{so-dump} released on March 2018 for data collection. Since we are mostly interested in the API classes discussed in the Q \& A texts, we adopt certain restrictions. First, we make sure that each question or answer contains a bit of code, \ie\ the thread is about coding. For this, we check the existence of \texttt{<code>} tags in their texts like the earlier studies \cite{low,explanation,surfclipse,qclassification}. Second, to ensure high quality content, we chose only such Q \& A threads where the answer was accepted as solution by the person who submitted the question \cite{tsc2016,saner2016masud}.
Once the Q \& A threads are collected,
 we perform standard natural language preprocessing (\ie\ removal of stop words, punctuation marks and programming keywords, token splitting) on each thread, and normalize their contents. Given the controversial evidence on the impact of stemming on source code \cite{stemming}, we avoid stemming on these threads given that they contain code segments. Our corpus is then indexed using \emph{Lucene}, a widely used search engine by the literature \cite{trconfig,refoqus,saner2017masud}, and later used for collecting feedbacks on a generic natural language query.

\textbf{Pseudo-Relevance Feedback (PRF) on the NL Query:} 
\citet{tsc2016} first employ Stack Overflow in collecting pseudo-relevance feedback on a given query. Their idea was to extract software-specific words relevant to a given query, 
and then to use them for query reformulation.
Similarly, we also collect pseudo-relevance feedback on the query using Stack Overflow. We first normalize a natural language query using standard natural language preprocessing (\ie\ stopword removal, token splitting), and then use it to retrieve Top-M (\eg\ $M=35$, check RQ$_1$ for detailed justification)
Q \& A threads from the above corpus with \emph{Lucene} search engine (\ie\ Fig. \ref{fig:sysdiag}, Steps 2--4, Algorithm \ref{algo}, Lines 4--8).  
The baseline idea is to extract appropriate API classes from them using appropriate selection methods \cite{concept-ase}, and then, to use them for query reformulation. 
We thus extract the program elements (\eg\ code segments, API classes) from each of the threads by analysing their HTML contents. We use \emph{Jsoup} \cite{jsoup}, a Java library for the HTML scraping.
We also develop \emph{two} separate sets of code segments from the questions and answers of the feedback threads. Then we use \emph{two} widely used term-weighting methods --\emph{TF-IDF} and \emph{PageRank}-- for collecting candidate API classes from them.   

\textbf{API Class Weight Estimation with TF-IDF:}
Existing studies \cite{refoqus,tsc2016,gayg} often apply Rocchio's method \cite{rocchio} for query reformulation where they use TF-IDF to select appropriate expansion terms. Similarly, we adopt TF-IDF for selecting potential reformulation candidates from the code segments that were collected above. In particular, we extract all API classes from each code segment (\ie\ feedback document) with the help of island parsing (\ie\ uses regular expressions) \cite{peter}, and then determine their relative weight (\ie\ Fig. \ref{fig:sysdiag}, Step 5, Algorithm \ref{algo}, Lines 11--12) as follows:
\begin{equation*}\label{eq:tfidf}
\setlength\abovedisplayskip{2pt}
\setlength\belowdisplayskip{2pt}
TF-IDF(A_i)=(1+log(TF_{A_i}))\times log(1+\frac{N}{DF_{A_i}})
\end{equation*} 
Here $TF_{A_i}$ refers to total occurrence frequency of an API class $A_i$ in the collected code segments, $N$ refers to total Q \& A threads in the corpus, and $DF_{A_i}$ is the number of threads that mentioned API class $A_i$ in their texts.

\textbf{API Class Weight Estimation with PageRank:} Semantics of a term are often determined by its contexts, \ie\ surrounding terms \cite{wembedding,wordsim}. Hence, inter-dependence of terms is an important factor in the estimation of term weight. However, TF-IDF assumes term independence (\ie\ ignores term contexts) in the weight estimation.  Hence, it might fail to identify highly important but not so frequent terms from a body of texts \cite{ase2017amasud,rada}. We thus employ another term weighting method that considers dependencies among terms in the weight estimation. In particular, we apply PageRank algorithm \cite{pagerank,rada} to the relevance feedback documents, \ie\ relevant code segments, and identify the important API classes as follows:


\begin{figure}[!t]
	\centering
	\begin{tikzpicture}[scale=.70, auto,swap]
	\begin{pgfonlayer}{bg}    
	
	\node [inner sep=1pt]  at (1.6,-1.2) [ellipse,draw, fill=yellow!20] (cc) {ColorConvertOp};
	\node [inner sep=1pt] at (-1.2,0) [ellipse,draw, fill=yellow!20] (cs) {ColorSpace};
	\node [inner sep=1pt] at (-1.2,1) [ellipse,draw, fill=gray!20] (file) {File};

	\node [inner sep=1pt] at (1.6,1.3) [ellipse,draw, fill=gray!20] (iio) {ImageIO};
	\node [inner sep=1pt] at (3.8,0) [ellipse,draw, fill=yellow!20] (bi) {BufferedImage};
	\node [inner sep=1pt] at (4,1) [ellipse,draw, fill=gray!20] (url) {URL};
	
	\draw[<->,thin] (bi) -- (iio);
	\draw[<->,thin] (iio) -- (url);
	\draw[<->,thin] (url) -- (bi);
	\draw[<->,thin] (bi) -- (cc);	
	\draw[<->,thin] (cc) -- (cs);
	\draw[<->,thin] (cs) -- (iio);
	\draw[<->,thin] (iio) -- (file);

	\end{pgfonlayer}
	
	\end{tikzpicture}
	\vspace{-.2cm}
	\caption{API co-occurrence graph for code segment in Listing \ref{lst:example}}
	\label{fig:token-graph}
	\vspace{-.6cm}
\end{figure}
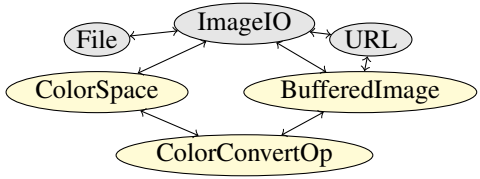

\textit{Construction of API Co-occurrence Graph:} Since PageRank algorithm operates on a graph-based structure, we transform pseudo-relevance feedback documents into a graph of API classes (\ie\ Fig. \ref{fig:sysdiag}, Step 6, Algorithm \ref{algo}, Line 13). In particular, we extract all API classes from each code segment using island parsing \cite{peter}, and then develop an ordered list by preserving their initialization order in the code. For example, the code snippet in Listing \ref{lst:example} is converted into a list of six API classes. 
Co-occurrences of items in a certain context has long been considered as an indication of relatedness among the items \cite{wordsim,rada}.
We thus capture the immediate co-occurrences of API classes in the above list, consider such co-occurrences as connecting edges, and then develop an API co-occurrence graph (\eg\ Fig. \ref{fig:token-graph}). 
We repeat the same step for each of the code segments, and update the connectivities in the graph. 
We develop one graph for the code segments from questions and another graph for the code segments from answers which were returned as a part of the pseudo-relevance feedback.


 
\textit{API Class Rank Calculation:} PageRank has been widely used for web link analysis \cite{pagerank} and term weighting in Information Retrieval domain \cite{rada}. It applies the underlying mechanism of recommendation or voting for determining importance of an item (\eg\ web page, term) \cite{saner2017masud}.
That is, PageRank considers a node as important only if it is recommended (\ie\ connected to) by other important nodes in the graph. 
The same idea has been widely used for separating legitimate pages from spam pages \cite{spam-pr}. 
Similarly, in our problem context, if an API class co-occurs with other important API classes across multiple code segments that are relevant to a programming task, then this API class is also considered to be important for the task. 
We apply PageRank algorithm on each of the two graphs (\ie\ Fig. \ref{fig:sysdiag}, Step 6, Algorithm \ref{algo}, Line 14), and determine the importance $ACR(v_i)$ (\ie\ API Class Rank) of each node $v_i$ by recursively applying the following equation: 
\begin{equation*}\label{eq:apirank}
\setlength\abovedisplayskip{2pt}
\setlength\belowdisplayskip{2pt}
ACR(v_{i})=(1-\phi)+\phi\sum_{j\epsilon In(v_{i})}\frac{ACR(v_{j})}{|Out(v_{j})|}~~ (0 \le \phi \le1)
\end{equation*} 
\noindent
Here, $In(v_i)$ refers to nodes providing inbound links (\ie\ votes) to node $v_i$ whereas $Out(v_i)$ refers to nodes that $v_i$ is connected to through outbound links, and $\phi$ is the damping factor. In the context of world wide web, \citet{pagerank} considered $\phi$ as the probability of visiting a web page
and $1-\phi$ as the probability of jumping off the page by a random surfer. We use a value $\phi=0.85$ for our work like the previous studies \cite{pagerank,rada,ase2017amasud}. We initialize each node with a score of 0.25, and run an \emph{iterative version} of PageRank on the graph. The algorithm pulls out weights from the surrounding nodes recursively, and updates the weight of a target node. This recursive process continues until the scores of the nodes converge below a certain threshold (\eg\ 0.0001 \cite{rada}) or total iteration count reaches the maximum (\eg\ 100 \cite{rada}). Once the computation is over, each node (\ie\ API class) is left with a score which is considered as a numerical proxy to its relative importance among all nodes. 

\textbf{Selection of Candidate API Classes:} 
Once two weights --TF-IDF and PageRank-- of each of the potential candidates are calculated, we rank the candidates according to their weights. Then we
select Top-N (\eg\ $N=16$, check RQ$_1$ for justification) API classes from each of the four lists (\ie\ two lists for each term weight, Fig. \ref{fig:sysdiag}, Step 7, Algorithm \ref{algo}, Lines 9--16). In Stack Overflow Q \& A site, a question often describes a programming problem (or a task) whereas the answer offers a solution. Thus, API classes involved with the problem and API classes forming the solution should be treated differently for identifying the \emph{relevant} and \emph{specific} API classes for the task. We leverage this inherent differences of context and semantics between questions and answers, and treat their code segments separately unlike the earlier study of \citet{tsc2016} that overlooks such differences. 


\begin{algorithm}[!tb]
	\caption{Query Reformulation using Relevant API Classes}
	\label{algo}
	\small
	\begin{algorithmic}[1]
		\Procedure{NLP2API}{$Q$}\Comment{$Q$: natural language query}
		\State $R \gets$ \{\}\Comment{$R$: Relevant API classes}
		\State $C \gets$developQ\&ACorpus($SODump$)\Comment{$C$: SO corpus}
		\State $Q_{pp} \gets$preprocess($Q$)
		\LineComment{collecting pseudo-relevance feedback}
		\State $PRF \gets $getPRF($Q_{pp}$, $C$)
		\State $PRF_{Q} \gets$ getQuestionCodeSegments($PRF$)
		\State $PRF_{A} \gets$ getAnswerCodeSegments($PRF$)
		\LineComment{collecting candidate API list}
		\For{PRF $prf$ $\in$ $\{PRF_{Q},PRF_{A}\}$}
		\State $TW \gets$calculateTFIDF($prf, C$)
		\State $WC[prf] \gets$getTopKWeightedClasses($TW$)
		\State $G \gets$developAPICo-occurrenceGraph($prf$)
		\State $ACR \gets$calculateAPIClassRank($G$)
		\State $RC[prf] \gets$getTopKRankedClasses($ACR$)
		\EndFor
		\LineComment{training the fastText model}
		\State $M_{ft} \gets$getFastTextModel(preprocess($SODump$))
		\LineComment{API relevance estimation}		
		\State $A \gets$getAllCandidateAPIClasses($RC \cup WC$)		
		\For{CandidateAPIClass $A_i$ $\in$ $A$}	
		\LineComment{calculate Borda score}			
		\State $S_B[A_i] \gets$ getBordaScore($A_i,RC,WC$)
		\LineComment{semantic relevance between API class and query}		
		\State $S_P[A_i] \gets$ getQuery-APIProximity($A_i, Q_{pp},M_{ft}$)
		\State $R[A_i].score\gets$ $S_B[A_i]$ + $S_P[A_i]$
		\EndFor
		\LineComment{ranking of the API classes}
		\State $rankedClasses \gets$ sortByFinalScore($R$)
		\LineComment{reformulation of the initial query}
		\State \textbf{return}  $Q_{pp} + rankedClasses$ 
		\EndProcedure
	\end{algorithmic}
\end{algorithm}
\setlength{\textfloatsep}{2pt}     
\subsection{Borda Score Calculation}\label{sec:borda}
Borda count is a widely used election method where the voters sort their political candidates on a scale of preference \cite{borda1,borda2}. In the context of Software Engineering, \citet{holmes} first apply Borda count to recommend relevant code examples for the code under development in the IDE. They apply this method to six ranked list of code examples collected using six structural heuristics, and then suggest 
the most frequent examples across these lists as the most relevant ones. 
Similarly, we apply this method to our four candidate API lists (\ie\ Fig. \ref{fig:sysdiag}, Step 8, Algorithm \ref{algo}, Lines 22--23) where each of the API classes are ranked based on their importance estimates (\eg\ TF-IDF, API Class Rank). We calculate Borda score $S_B$ for each of the API classes ($\forall A_i\in A$) from the these ranked candidate lists--$WRC=\{WC_Q, WC_A, RC_Q, RC_A\}$--as follows:  
\begin{equation*}\label{eq:apirank}
\setlength\abovedisplayskip{2pt}
\setlength\belowdisplayskip{2pt}
S_B(A_i\in A)=\sum_{RL_j\in WRC} 1-\frac{rank(A_i,RL_j)}{|RL_j|}   
\end{equation*}
\noindent
Here, $A$ refers to the set of all API classes extracted from the ranked candidate lists --$WRC$, $|RL_j|$ denotes each list size, and $rank(A_i,RL_j)$ returns the rank of class $A_i$ in the ranked list. Thus, an API class that occurs at the top positions in multiple candidate lists is likely to be more important for a target  programming task than the ones that either occurs at the lower positions or does not occur in multiple lists.       

\subsection{Query-API Semantic Proximity Analysis}\label{sec:w2vec}
Pseudo-relevance feedback, PageRank (Section \ref{sec:candidate}) and Borda count (Section \ref{sec:borda})  
analyse local contexts of the query keywords within a set of tentatively relevant documents (\ie\ Q \& A threads) and then extract candidate API classes for query reformulation. Although local context analysis is useful, existing studies report that such analysis alone might cause topic drift from the original query \cite{li2016,qsurvey}. We thus further analyse global contexts of the query keywords, and determine the semantic proximity between the given natural language query and the candidate API classes as follows:


\textbf{Word2Vec Model Development:} \citet{w2vec} and colleagues propose a neural network based tool--\textit{word2vec}--for learning word embeddings from an ultra-large body of texts where they employ continuous bag of words (CBOW) and skip-gram models. While other studies attempt to define context of a word using co-occurrence frequencies or TF-IDF \cite{wordsim, saner2016masud,irmarcus}, they offer a probabilistic representation of the context. In particular, they learn \emph{word embeddings} (Section \ref{sec:bg}) for each of the words from the corpus, and map each word to a point in the semantic space so that semantically similar words appear in the close proximity.   
We leverage this notion of \emph{semantic proximity}, and determine the relevance 
of a candidate API class to the given query.
It should be noted that such proximity measure could be an effective tool to overcome the \emph{vocabulary mismatch issues} \cite{vocaprob}. We thus develop a \textit{word2vec} model where 1.3 million programming questions and answers (\ie\ 656,538 Q \& A pairs, collected in Section \ref{sec:candidate}) are employed as the corpus. We normalize each question and answer using standard natural language preprocessing, and learn the word embeddings (Fig. \ref{fig:sysdiag}, Step 1b, 1c, 1d, Algorithm \ref{algo}, Lines 17--18) using skip-gram model.
For our learning, we use \textit{fastText} \cite{fasttext}, an improved version of \textit{word2vec} that incorporates sub-word information into the model.
We performed the learning offline and it took about one hour.
It should be noted that our model is learned using default parameters (\eg\ output vector size = 100, context window size = 5, minimum word occurrences = 5) provided by the tool.

\textbf{Semantic Relevance Estimation:} While a given query contains multiple keywords, a candidate API class might not be semantically close to all of them. We thus capture the maximum proximity estimate between an API class and any of the query keywords as the relevance estimate of the class. In particular, we collect word embeddings (\ie\ a vector of 100 real valued estimates of the contexts) of each candidate API class $A_i\in A$ and each keyword $q\in Q$, and determine their semantic proximity $S_P$ using \emph{cosine similarity} (\ie\ Fig. \ref{fig:sysdiag}, Step 9, Algorithm \ref{algo}, Lines 24--25) as follows:
\begin{equation*}\label{eq:semantic}
\setlength\abovedisplayskip{2pt}
\setlength\belowdisplayskip{2pt}
S_P(A_i\in A)= \{f(A_i,q)~|~f(A_i,q)>f(A_i,q_0)\forall q_0\in Q \}
\end{equation*}
\begin{equation*}\label{eq:semantic}
\setlength\abovedisplayskip{0pt}
\setlength\belowdisplayskip{0pt}
f(A_i, q)=cosine(fastText(A_i), fastText(q))
\end{equation*}
\noindent
Here $fastText(.)$ returns the learned word embeddings of either a query keyword or an API class, and $f(A_i,q)$ returns the \emph{cosine similarity} between their word embeddings.
We use \texttt{print-word-vectors} option of \textit{fastText}, and collect the word embeddings from our learned model on Stack Overflow. 

\subsection{API Class Relevance Ranking \& Query Reformulation}\label{sec:ranking}
Once Borda score $S_B$ and semantic proximity score $S_P$ are calculated, we normalize both scores between 0 and 1, and then sum them up using a \emph{linear combination} (\ie\ Line 26, Algorithm \ref{algo}) for each of the candidate API classes. While fine tuned relative weight estimation for these two scores could have been a better approach, we keep that as a part of future work. Besides, equal weights also reported pretty good results (\eg\ 82\% Top-10 accuracy) according to our investigation.
The API classes are then ranked according to their final scores, and Top-K (\eg\ $K=10$) classes are suggested as the relevant classes for the programming task stated as a generic query (\ie\ Fig. \ref{fig:sysdiag}, Steps 10--12, Algorithm \ref{algo}, Lines 19--29). These API classes are then appended to the given query as reformulations \cite{refoqus} (\ie\ Fig. \ref{fig:sysdiag}, Steps 13, Algorithm \ref{algo}, Lines 30--31). Table \ref{table:exampleref} shows our reformulated query for the showcase natural language query using the suggested API classes.

\section{Experiment}\label{sec:experiment}
We conduct experiments with 310 code search queries   
randomly collected from four popular programming tutorial sites, and evaluate our query reformulation technique. We choose five appropriate performance metrics from the literature, and evaluate two aspects of our provided supports-(1) relevant API class suggestion and (2) query reformulation. Our technique is also validated against three state-of-the-art techniques \cite{saner2016masud,tsc2016,cocabu} and three popular code/web search engines including Google. We thus answer five research question using our experiments as follows: 
\begin{itemize}[noitemsep,topsep=0pt]
\item \textbf{RQ$\mathbf{_1}$:} How does NLP2API perform in recommending relevant API classes for a given query?
How do different parameters and thresholds influence the performance?
\item \textbf{RQ$\mathbf{_2}$:} Can NLP2API outperform the state-of-the-art technique on relevant API class suggestion for a query?
\item \textbf{RQ$\mathbf{_3}$:} Can the reformulated queries of NLP2API outperform the baseline natural language queries?
\item \textbf{RQ$\mathbf{_4}$:} Can NLP2API outperform the state-of-the-art technique on query reformulation that uses crowdsourced knowledge from Stack Overflow?
\item \textbf{RQ$\mathbf{_5}$:} Can NLP2API significantly improve the results provided by state-of-the-art code or web search engines? 
\end{itemize}

\subsection{Experimental Dataset}\label{sec:expds}
\textbf{Dataset Collection:} We collect 310 code search queries from four popular programming tutorial sites--KodeJava \cite{kodejava}, Java2s \cite{java2s}, CodeJava \cite{codejava} and JavaDB \cite{javadb}--for our experiments. While 150 of these queries were taken from a publicly available dataset \cite{saner2016masud}, we attempted to extend the dataset by adding 200 more queries. However, after removing the duplicates and near duplicates, we ended up with 160 queries. Thus, our dataset contains a total of 310 (\ie\ 150 old + 160 new)  search queries.  
Each of these sites above discusses hundreds of programming tasks as Q \& A threads where each thread generally contains (1) a question title, (2) a solution (\ie\ code), and (3) a prose explaining the code succinctly. The question title (\eg\ \emph{``How do I decompress a GZip file in Java?"} \cite{qtitle}) generally comprises of a few important keywords and often \emph{resembles} a real life search query. We thus use these titles from tutorial sites as code search queries in our experiments, as were also used by the earlier studies \cite{saner2016masud,conngraph}.

\textbf{Ground Truth Preparation:} The prose that explains code in the tutorial sites above often includes one or more API classes from the code (\eg\ \texttt{GZipInputStream}, \texttt{FileOutputStream}). Since these API classes are chosen to explain the code that implements a programming task, they are generally relevant and specific to the task.
We thus consider these \emph{relevant} and \emph{specific} API classes as the \emph{ground truth} for the corresponding question title (\ie\ our search query) \cite{saner2016masud}. We develop a \emph{ground truth API set} to evaluate the performance of our technique in the API class suggestion. We also collect the code segments from each of the 310 Q \& A threads from the tutorial sites above as the \emph{ground truth code segments}, and use them to evaluate the query reformulation performance (\ie\ in terms of code retrieval) of our technique. Given that these API classes and code segments are publicly available online and were consulted by thousands of technical users over the years, subjectivity associated with their relevance to the corresponding tasks (\ie\ our selected queries) is minimized \cite{conngraph}. Our dataset preparation step took \textbf{$\approx$ 25} man hours. 

\textbf{Replication Package:} Our dataset, working prototype and other materials are \emph{accepted} for publication \cite{icsme2018masudb}. They are publicly available \cite{nlp2api} for replication and third party reuse.   

\subsection{Performance Metrics}\label{sec:pmetric}
We choose five performance metrics that were widely adopted by relevant literature \cite{portfolio,conngraph,feature,tsc2016,saner2016masud,saner2017masud}, for the evaluation and validation of our technique as follows:

\textbf{Top-K Accuracy / Hit@K}: It is the percentage of search queries for each of which at least one item (\eg\ API class) from the \emph{ground truth} is returned within the Top-K results.  

\textbf{Mean Reciprocal Rank@K (MRR@K)}: Reciprocal Rank@K is defined as the multiplicative inverse of the rank of first relevant item
(\eg\ API class from ground truth) in the Top-K results returned by a technique. Mean Reciprocal Rank@K (MRR@K) averages such measures for all queries. 

\textbf{Mean Average Precision@K (MAP@K)}: Precision@K is the precision calculated at the occurrence of K$^{th}$ item in the ranked list. Average Precision@K (AP@K) averages the precision@K for all relevant items (\eg\ API class from ground truth) within the Top-K results for a search query. Mean Average Precision@K is the mean of Average Precision@K for all queries from the dataset. 

\textbf{Mean Recall@K (MR@K)}: Recall@K is defined as the percentage of ground truth items (\eg\ API classes) that are correctly recommended for a query in the Top-K results by a technique. Mean Recall@K (MR@K) averages such measures for all queries from the dataset. 

\textbf{Query Effectiveness (QE):} It is defined as the rank of first correct item (\ie\ ground truth code segment) in the result list returned by a query. The measure is an approximation of the developer's effort in locating the first code segment relevant to a given query. Thus, the lower the effectiveness measure is, the more effective  the query is \cite{saner2017masud,trconfig}. We use this measure to 
evaluate the improvement of a query through reformulations offered by a technique. 


\begin{figure}[!tb]
	\centering
	\includegraphics[width=3.2in]{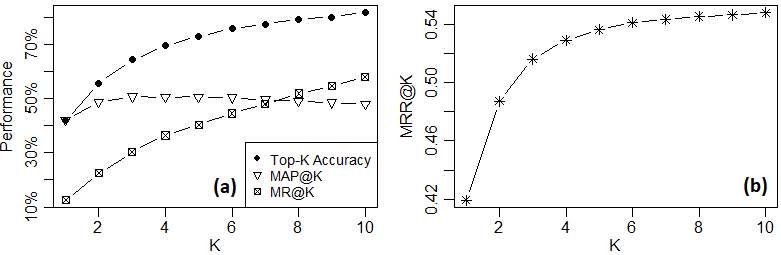}
	\vspace{-.3cm}
	\caption{Performance of NLP2API in API class suggestion for various Top-K results}
	\vspace{-.3cm}
	\label{fig:topk}
\end{figure}
  
\begin{table}[!tb]
	\centering
	\caption{Performance of NLP2API in Relevant API Suggestion}\label{table:expresult}
	\vspace{-.2cm}
	\resizebox{3.3in}{!}{%
		\begin{threeparttable}
			\begin{tabular}{l|c|c|c|c}
				\hline
				\textbf{Performance Metric} & \textbf{Top-1} & \textbf{Top-3} & \textbf{Top-5}& \textbf{Top-10}\\
				\hline
				\hline
				Top-K Accuracy & 41.94\% & 64.19\% & \textbf{72.90}\% & \textbf{81.61}\% \\
				\hline
				Mean Reciprocal Rank@K & 0.42 & 0.52 & \textbf{0.54} & \textbf{0.55} \\
				\hline
				Mean Average Precision@K & 41.94\% & \textbf{50.62}\% & \textbf{50.56}\% & 47.85\%\\
				\hline
				Mean Recall@K & 12.53\% & 30.17\% & \textbf{40.28}\% & \textbf{57.87}\%\\
				\hline
			\end{tabular}
		\centering
		\textbf{Top-K} = Performance measures for Top-K suggestions
		\end{threeparttable}
	}
\end{table}

\subsection{Evaluation of NLP2API: Relevant API Class Suggestion}\label{sec:eval-api}
We first evaluate the performance of our technique in the relevant API class suggestion for a generic code search query. We make use of 310 code search queries (Section \ref{sec:expds}) and four performance metrics (Section \ref{sec:pmetric}) for this experiment.
We collect Top-K (\eg\ $K$=10) API classes suggested for each query, compare them with the \emph{ground truth API classes}, and then determine our API suggestion performance. In this section, we also answer RQ$_1$ and RQ$_2$ as follows:  


\textbf{Answering RQ$\mathbf{_1}$--Relevant API Class Suggestion:} From Table \ref{table:expresult}, we see that our technique returns relevant API classes for 73\% of the queries with 51\% mean average precision and 40\% recall when only Top-5 results are considered. That is, half of the suggested classes come from the \emph{ground truth}, and our approach succeeds for seven out of 10 queries. More importantly, it achieves a mean reciprocal rank of 0.54. That means, on average, the first relevant API class can be found at the second position of the result list. Such classes can also be found at the first position for 42\% of the queries.
All these statistics are highly promising according to relevant literature \cite{feature,saner2016masud}.
Fig. \ref{fig:topk} further demonstrates our performance measures for Top-1 to Top-10 results. We see that accuracy, recall and reciprocal rank measures increase
monotonically which are expected. Interestingly, the precision measure shows an almost steady behaviour. That means, as more results were collected, our technique was able to \emph{filter out} the \emph{false positives} which demonstrates its high potential for API suggestion. 




   
 \begin{figure}[!tb]
 	\centering
 	\includegraphics[width=3.20in]{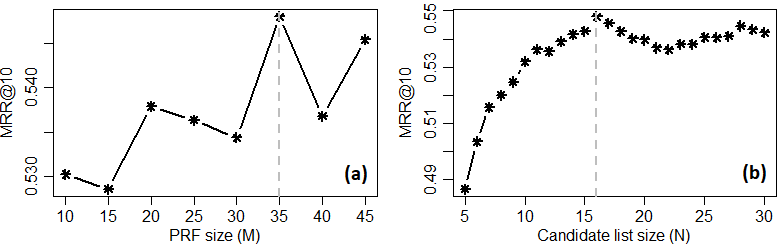}
 	\vspace{-.3cm}
 	\caption{Impact of (a) PRF size (M), and (b) Candidate API list size (N) on relevant API class suggestion from Stack Overflow}
 	\label{fig:mnsize}
 \end{figure} 

\textbf{Impact of Pseudo-Relevance Feedback Size (M) and Candidate API List Size (N):} We investigate how different sizes of pseudo-relevance feedback (\ie\ number of Q \& A threads retrieved from Stack Overflow by the given query) and candidate API list (\ie\ detailed in Section \ref{sec:candidate}) affect the performance of our technique. We conduct experiments using 10--45 feedback Q \& A threads and 5--30 candidate API classes. We found that these parameters improved accuracy and recall measures monotonically (\ie\ as expected) but affected precision measures in an irregular fashion (\ie\ not monotonic). However, we found an interesting pattern with mean reciprocal rank. From Fig. \ref{fig:mnsize}, we see that mean reciprocal rank@10 of our technique reaches 
the maximum when (a) pseudo-relevance feedback size, $M$ is 35 and (b) candidate API list size, $N$ is 16. 
We thus adopt these thresholds, 
\ie\ $M=35$ and $N=16$, in our technique for the experiments.

\begin{figure}[!tb]
	\centering
	\includegraphics[width=3.25in]{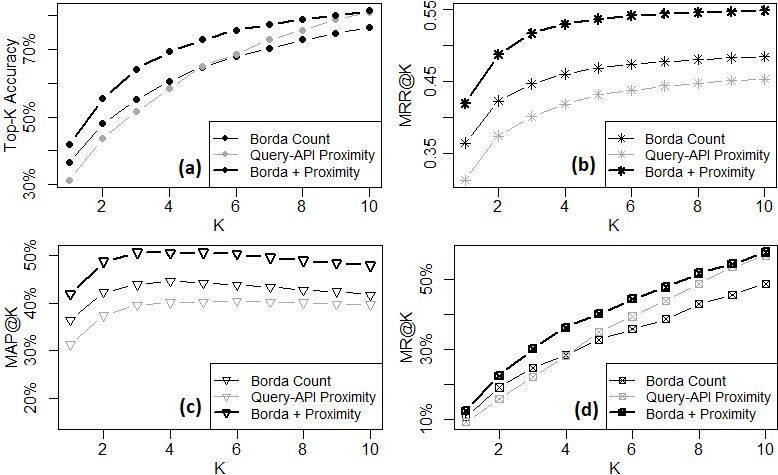}
	\vspace{-.4cm}
	\caption{Comparison between Borda count and Query-API proximity in estimating API relevance using (a) accuracy, (b) reciprocal rank, (c) precision, and (d) recall}
	\label{fig:borda-w2vec}
	\vspace{-.5cm}
\end{figure}  

\textbf{Borda Count vs. Query-API Class Proximity as API Relevance Estimate:} Once candidate API classes are selected (Section \ref{sec:candidate}), we employ two proxies (Sections \ref{sec:borda}, \ref{sec:w2vec}) for estimating the relevance of an API class to the NL query. We compare the appropriateness of these proxies-- \emph{Borda Count} and \emph{Query-API Proximity}-- in capturing the API class relevance, and report our findings in Fig. \ref{fig:borda-w2vec}.
We see that Borda Count is more effective than Query-API Proximity in capturing the relevance of an API class to a given query. However, the proximity demonstrates its potential especially with accuracy and recall measures.
More interestingly, combination of these two proxies ensures the best performance of our technique in all four metrics. Non-parametric statistical tests also report that performances with Borda+Proximity are significantly higher than those with either Borda Count (\ie\ all \emph{p-values$<$0.05}, $0.34\le\Delta\le 0.82$ (\emph{large})) or Query-API Semantic Proximity (\ie\ all \emph{p-values$<$0.05}, $0.20\le\Delta\le 0.90$ (\emph{large})).  
 
We also investigate the parameters of \emph{fastText} \cite{fasttext} that were used to determine the query-API proximity. Although we experimented using various custom parameters, we did not see any significant performance gain over the default parameters. Besides, increased thresholds (\eg\ context window size, output vector size) could be computationally costly. We thus adopt the default settings of \emph{fastText} in this work.
\vspace{-.2cm}
\FrameSep.3em
\begin{framed}
	\noindent
	Our technique provides the first relevant API class at the \emph{second} position, $\approx$ \textbf{50}\% of our suggested classes are true positive, and the technique succeeds \emph{eight} out of 10 times (\ie\ \textbf{82}\% Top-10 accuracy). Besides, our adopted parameters and thresholds (\eg\ $M$, $N$) are justified. 
\end{framed}
\vspace{-.2cm}


 \begin{table}[!tb]
 	\centering
 	\caption{Comparison with the State-of-the-art in API Class Suggestion}\label{table:comparison-api}
 	\vspace{-.2cm}
 	\resizebox{3.3in}{!}{%
 		\begin{threeparttable}
 			\begin{tabular}{l|l|c|c|c|c}
 				\hline
 				\textbf{Technique} & \textbf{Metric} & \textbf{Top-1} & \textbf{Top-3} & \textbf{Top-5} & \textbf{Top-10}\\
 				\hline
 				\hline
 				\multirow{4}{*}{RACK \cite{saner2016masud}} & Top-K Accuracy & 20.97\% & 52.90\% & 64.19\% & 77.10\%\\ 
 				\hhline{~-----}
 				& MRR@K & 0.21 & 0.35 & 0.37 & 0.39\\				
 				\hhline{~-----}
 				& MAP@K  & 20.97\% & 34.76\% & 36.76\% & 36.38\%\\
 				\hhline{~-----}
 				& MR@K & 6.25\% & 20.81\% & 28.06\% & 39.22\%\\
 				\hline
 				\hline
 				\multirow{2}{*}{}& Top-K Accuracy & 41.94\% & 64.19\% & \textbf{72.90}\% & \textbf{81.61}\% \\
 				\hhline{~-----}
 				 \textbf{NLP2API} & MRR@K &  0.42 & 0.52 & \textbf{0.54} & \textbf{0.55} \\
 				\hhline{~-----}
 				(Proposed) & MAP@K & 41.94\% & \textbf{50.62}\% & \textbf{50.56}\% & 47.85\% \\
 				\hhline{~-----}
 				& MR@K & 12.53\% & 30.17\% & \textbf{40.28}\% & \textbf{57.87}\% \\
 				\hline
 			\end{tabular}
 			\centering
 			\textbf{Top-K} = Performance measures for Top-K suggestions
 		\end{threeparttable}
 	}
 	\vspace{-.1cm}
 \end{table}
 
 
\textbf{Answering RQ$\mathbf{_2}$-- Comparison with Existing Studies on Relevant API Class Suggestion:}
We compare our technique with the state-of-the-art approach -- RACK \cite{saner2016masud} -- on API class suggestion for a natural language query. \citet{saner2016masud} employ two heuristics-- Keyword-API Co-occurrence (KAC) and Keyword-Keyword Coherence (KKC)--for suggesting relevant API classes from Q \& A threads of Stack Overflow for a given query. Their approach outperformed earlier approaches \cite{feature,conngraph} which made it the state-of-the-art in relevant API class suggestion. We collect the authors' implementation of RACK from corresponding web portal, ran the tool as is on our dataset, and then extract the evaluation results. 


From Table \ref{table:comparison-api}, we see that our technique--NLP2API-- outperforms RACK especially in precision, recall and reciprocal rank. 
It should be noted that our reported performance measures for RACK are pretty close to the authors' reported measures \cite{saner2016masud}, which indicates a \emph{fair comparison}. We see that RACK recommends API classes correctly for 64\% of the queries with 37\% precision, 28\% recall and a reciprocal rank of 0.37 when Top-5 results are considered. On the contrary, our technique recommends correctly for 73\% of the queries with 51\% precision, 40\% recall and a promising reciprocal rank of 0.54 
in the same context. These are 14\%, 38\%, 44\% and 46\% improvement respectively over the state-of-the-art performance measures. 
Statistical tests for various Top-K results (\ie\ 1$\le$K$\le$10) also reported significance (\ie\ all \emph{p-values}$\le$0.05) of our technique over the state-of-the-art with large effect sizes (\ie \ $0.39\le\Delta\le 0.90$). 
\vspace{-.2cm}
\FrameSep.3em
\begin{framed}
\noindent
Our technique outperforms the state-of-the-art approach on relevant API class suggestion, and it suggests relevant API classes with \textbf{38}\% higher precision and \textbf{46}\% higher reciprocal rank than those of the state-of-the-art. 
\end{framed}  

\subsection{Evaluation of NLP2API: Query Reformulation} \label{sec:eval-qr}
Although our approach outperforms the state-of-the-art on relevant API class suggestion, we further apply the suggested API classes to query reformulations. Then we demonstrate the potential of our reformulated queries for improving the code snippet search. In this section, we also answer RQ$_3$, RQ$_4$ and RQ$_5$ using our experiments as follows:  
    
\begin{table}
	\centering
	\caption{Impact of Reformulations on Generic NL Queries}\label{table:baseline}
	\vspace{-.2cm}
	\resizebox{3.15in}{!}{%
		\begin{threeparttable}
			\begin{tabular}{l|c|c|c|c}
				\hline
				\textbf{Reformulation}& \textbf{RL} & \textbf{Improved/MRD} & \textbf{Worsened/MRD} & \textbf{Preserved}\\
				\hline
				\hline
				\multirow{2}{*}{NLP2API$_B$} & 05 & 43.23\%/-245 & 31.29\%/+54  & 25.48\% \\
				\hhline{~----}
				& 10 & \textbf{48.07}\%/-223 & 26.13\%/+65  & 25.81\% \\
				\hline
				NLP2API$_P$ & 10 & 40.97\%/-148 & 30.97\%/+44  & 28.06\% \\
				\hline
				\multirow{3}{*}{\textbf{NLP2API}} & 05 & 40.00\%/-159 & 27.74\%/+54 & 32.26\% \\
				\hhline{~----}
				& 10 & \textbf{48.07}\%/-209 & 25.16\%/+45 & \textbf{26.77}\%\\
				\hhline{~----}
				& 15 & \textbf{49.03}\%/-217 & \textbf{22.26}\%/+46 & \textbf{28.71}\% \\
				\hline
			\end{tabular}
			\centering
			\textbf{MRD} = Mean Rank Difference between reformulated and given queries 
		\end{threeparttable}
	}
	\vspace{-.1cm}
\end{table}

 \begin{figure}[!tb]
 	\centering
 	\includegraphics[width=3.2in]{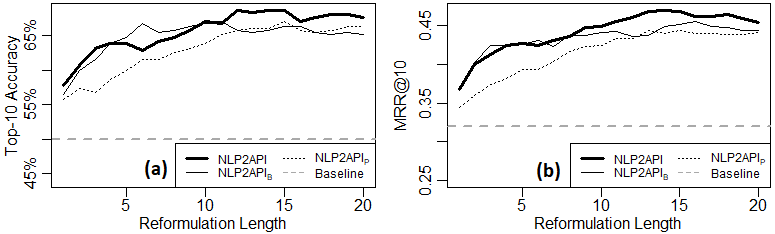}
 	\vspace{-.3cm}
 	\caption{Reformulated vs. baseline query using (a) Top-10 accuracy and (b) MRR@10}
 	\label{fig:baseline-improved}
 	\vspace{-.1cm}
 \end{figure}

\begin{table*}[!t]
	\centering
	\caption{Comparison of Query Effectiveness with Existing Query Reformulation Techniques }\label{table:comparison-qe}
	\vspace{-.2cm}
	\centering
	\resizebox{7in}{!}{%
		\begin{threeparttable}
			\begin{tabular}{l|c|c|c|c|c|c|c|c||c|c|c|c|c|c|c||c}
				\hline
				\multirow{2}{*}{\textbf{Technique}} &  \multirow{2}{*}{\textbf{\#QC}} &\multicolumn{7}{c||}{\textbf{Improvement}} & \multicolumn{7}{c||}{\textbf{Worsening}} & \textbf{Preserving}   \\
				\hhline{~~---------------}
				&   & \#Improved & Mean & Q1 & Q2 & Q3 & Min. & Max. & \#Worsened & Mean & Q1 & Q2 & Q3 & Min. & Max. & \#Preserved\\
				\hline
				\hline
				QECK \cite{tsc2016} & 310 & 72 (23.23\%) & 139 & 02 & 11 &  74 & 01 & 1,861 &  177 (57.10\%) & 131 & 11 & 35 & 163 & 02 & 1,259 & 61 (19.68\%) \\
				\hline
				\textbf{RACK} \cite{saner2016masud} & 310 & 105 (33.87\%) & 75 & 02 & 08 & 60 & 01 & 971 &	 147 (47.42\%) & 136 & 07 & 31 & 156 & 02 & 1,277 & 58 (18.71\%)\\
				\hline
				\textbf{CoCaBu} \cite{cocabu} & 310 & \textbf{113 (36.45\%)} & 191 & 02 & 14 & 103 & 01 & 2,607 & \textbf{131 (42.26\%)} & 102 & 06 & 24 & 91 & 02 & 1,567 & 66 (21.29\%) \\ 
				\hline
				\hline
				Baseline  & 310 & - & - & 07 & 25 & 145 & 02 & 1,460 & - & - & 01 & 03 & 15 & 01 & 582 & -\\
				\hline
				\textbf{NLP2API}  & 310 & \textbf{149 (48.07\%)} & \textbf{170} & \textbf{02} & \textbf{12} & \textbf{74} & \textbf{01} & 2,816 & \textbf{78 (25.16\%)} & \textbf{75} & \textbf{03} & \textbf{13} & \textbf{59} & \textbf{02} & \textbf{826} & \textbf{83 (26.77\%)}\\
				\hline
				\textbf{NLP2API}$_{max}$ & 310 & \textbf{152 (49.03\%)} & \textbf{172} & \textbf{02} & \textbf{10} & \textbf{61} & \textbf{01} & 2,926 & \textbf{69 (22.26\%)} & \textbf{73} & \textbf{03} & \textbf{11} & \textbf{70} & \textbf{02} & \textbf{786} & \textbf{89 (28.71\%)}\\
				\hline
			
			\end{tabular}
			\centering
			\textbf{Mean} = Mean rank of first correct results returned by the queries, \textbf{Q$_i$}=  $i^{th}$ quartile of all ranks considered 
		\end{threeparttable}
	}
	\vspace{-.5cm}
\end{table*}

  \textbf{Answering RQ$\mathbf{_3}$--Improvement of Natural Language Queries with the Suggested API Classes:} 
  We reformulate each of the generic natural language queries for code search using the API classes suggested by our technique. Then we investigate the performance of these reformulated queries using code search. We prepare a code corpus of 4,170 code segments where 310 segments are \emph{ground truth} code segments (Section \ref{sec:expds}) and 3,860 code segments were taken from a publicly available and curated dataset \cite{scam2014masud} based on hundreds of GitHub projects. We normalize these segments using standard natural language preprocessing (\ie\ stop and keyword removal, token splitting), and index them with \emph{Lucene}. 
  We then perform code search on this corpus, and contrast between generic natural language queries and our reformulated queries in terms of their Effectiveness and code retrieval performances.

  From Table \ref{table:baseline}, we see that our reformulations improve or preserve 
  75\% (\ie\ 48\% improvement and 27\% preserving) of the given queries. The improvement ratio reaches the maximum of 49\% with a reformulation length of 20.
  According to relevant literature \cite{refoqus,saner2017masud,trconfig}, such statistics are promising.       
  Fig. \ref{fig:baseline-improved} further demonstrates the impact of our reformulations on the baseline generic queries. 
  We see that the baseline natural language queries retrieve ground truth code segments with 50\% Top-10 accuracy (dashed line, Fig. \ref{fig:baseline-improved}-(a)) and 0.32 mean reciprocal rank (dashed line, Fig. \ref{fig:baseline-improved}-(b)). 
  On the contrary, our reformulated queries achieve a maximum of 69\% Top-10 accuracy with a reciprocal rank of 0.47 which are 37\% and 47\% higher respectively than the baseline. 
  Quantile analysis in Table \ref{table:comparison-qe} also shows that our provided result ranks are more promising than those of the baseline queries. 
  \vspace{-.2cm}
  \FrameSep.3em
  \begin{framed}
  \noindent
  Reformulations offered by our technique improve \textbf{49}\% of the generic natural language queries, and the reformulated queries achieve \textbf{37}\% higher accuracy and \textbf{47}\% higher reciprocal rank than those of the generic NL queries.  
  \end{framed}
\vspace{-.2cm}
  
 
\textbf{Answering RQ$_4$--Comparison with Existing Query Reformulation Techniques:} \citet{tsc2016} collect pseudo-relevance feedbacks from Stack Overflow on a given query and then apply Rocchio's method to expand the query. Their approach, QECK, outperformed earlier studies \cite{portfolio,pwordnet} on query reformulation targeting code search which made it the state-of-the-art. Another contemporary work, CoCaBu \cite{cocabu} applies Vector Space Model (VSM) in identifying appropriate program elements from Stack Overflow posts.    
To the best of our knowledge, these are the most recent and most closely related works to ours. Due to the unavailability of authors' prototype,  
we re-implement them ourselves using their best performing parameters (\eg\ PRF size = 5--10, reformulation length = 10), and then compare them with ours. We also compare with RACK \cite{saner2016masud} in the context of query reformulation due to its highly related nature.

Table \ref{table:comparison-qe} shows a quantile analysis of the result ranks provided by the existing techniques. If results are returned closer to the top of the list by a reformulated query than its baseline counterpart, we call it \emph{query improved} and vice versa as \emph{query worsened}.  
We see that CoCaBu and RACK perform relatively higher than QECK. CoCaBu improves 36\%
and worsens 42\% of the 310 baseline queries. 
On the contrary, our technique improves 48\% and worsens 25\% of the given queries which are 32\% higher and 40\% lower respectively than those of CoCaBu. Furthermore, according to the quantile analyses, the extents of our rank improvement over the baseline are comparatively higher than the extents of rank worsening which indicates a net benefit of the reformulation operations. 
\vspace{-.2cm}
\FrameSep.3em
\begin{framed}
	\noindent
	Our technique outperforms the state-of-the-art approaches on query reformulation, and it improves \textbf{32}\% more and worsens \textbf{40}\% less queries than those of the state-of-the-art.   
\end{framed}
\vspace{-.3cm}

\textbf{Answering RQ$_5$--Comparison with Existing Code/Web Search Engines:} Although our approach outperforms the state-of-the-art studies \cite{saner2016masud,tsc2016} on relevant API suggestion and query reformulation, we further compare with two popular web search engines -- \emph{Google, Stack Overflow native search} -- and one popular code search engine --\emph{GitHub code search}. Given the enormous and \emph{dynamic index database} and \emph{restrictions} on the \emph{query length} or \emph{type}, a full scale or direct comparison with these search engines is neither feasible nor fair. We thus investigate whether results returned by these contemporary search engines for generic queries could be significantly improved or not with the help of our reformulated queries. 

\textbf{Collection of Search Results and Establishment of Ground Truth:}
We first collect Top-30 results returned by each search engine for each of the 310 queries. For result collection, we make use of \emph{Google's custom search API} \cite{cse} and the native API endpoints provided by Stack Overflow and GitHub. Since our goal is to find relevant code snippets, we adopt a pragmatic approach in the establishment of ground truth for this experiment. In particular, we analyse those 30 results semi-automatically, look for \emph{ground truth code segments} (\ie\ collected in Section \ref{sec:expds}) in their contents, and then select Top-10 results  as \emph{ground truth search results} that contain either the ground truth code or highly similar code. It should be noted that ground truth code segments and our suggested API classes are taken from two different sources.

\begin{table}
	\centering
	\caption{Comparison with Popular Web/Code Search Engines}\label{table:comparison-se}
	\vspace{-.2cm}
	\resizebox{2.8in}{!}{%
		\begin{threeparttable}
			\begin{tabular}{l|c|c|c|c}
				\hline
				\textbf{Technique}  & \textbf{Hit@10} & \textbf{MAP@10} & \textbf{MRR@10} & \textbf{NDCG@10}\\
				\hline
				\hline
				Google  & 100.00\% & 65.50\% & 0.80 & 0.47 \\
				\hline
				\textbf{NLP2API}$_{Google}$  & 100.00\% & \textbf{76.73}\% & \textbf{0.83} & \textbf{0.61} \\
				\hline
				\hline
				Stack Overflow  & 90.65\% & 59.46\% & 0.67 & 0.40 \\  
				\hline
				\textbf{NLP2API}$_{SO}$  & 91.29\% & \textbf{79.95}\% & \textbf{0.87} & \textbf{0.67}\\
				\hline
				\hline
				GitHub  & 88.06\% & 53.06\% & 0.55 & 0.41 \\
				\hline
				\textbf{NLP2API}$_{GitHub}$ & 89.03\% & \textbf{70.69}\% & \textbf{0.78} & \textbf{0.59}\\
				\hline
			\end{tabular}
			\centering
			\textbf{NDCG}=Normalized Discounted Cumulative Gain \cite{ndcg2}
		\end{threeparttable}
	}
	\vspace{-.3cm}
\end{table}

\begin{figure}[!tb]
	\centering
	\includegraphics[width=3.25in]{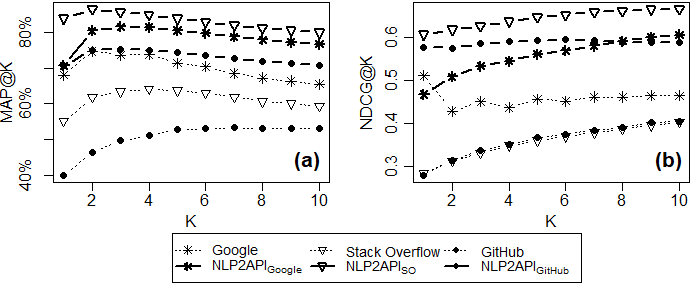}
	\vspace{-.2cm}
	\caption{Comparison between popular web/code search engines and NLP2API in relevant code segment retrieval using (a) MAP@K and (b) NDCG@K}
	\label{fig:compare-se}
\end{figure}

\textbf{Comparison between Initial Search Results and Re-ranked Results with Reformulated Queries:} While the search engines return results mostly for the natural language queries, we further re-rank the results with our reformulated queries (\ie\ generic search keywords + relevant API classes) using lexical similarity analysis (\eg\ cosine similarity \cite{scam2014masud}).
We then evaluate Top-10 results both by each search engine and by our re-ranking approach against the \emph{ground truth search results}, and demonstrate the potential of our reformulations.     

From Table \ref{table:comparison-se}, we see that the re-ranking approach that leverages our reformulated queries improves the initial search results returned by each of the engines. In particular, the performances are improved in terms of precision and discounted cumulative gain. For example, Google returns search results with 66\% precision and 0.47 NDCG when Top-10 results are considered. Our approach, NLP2API$_{Google}$, improves the ranking and achieves a MAP@10 of 77\% and a NDCG@10 of 0.61 which are 17\% and 30\% higher respectively. That is, although Google performs high as a \emph{general purpose} web search engine, it might always not be precise for \emph{code search} due to the lack of appropriate contexts. Our approach incorporates context into the search using relevant API names, and delivers more precise code search results.   
As shown in Table \ref{table:comparison-se} and Fig. \ref{fig:compare-se}, similar findings were also achieved against GitHub code search and Stack Overflow native search.
\vspace{-.6cm}
\FrameSep.3em
\begin{framed}
	\noindent
	Our technique improves upon the result ranking of all three popular search engines using its reformulated queries. It achieves \textbf{17}\% higher precision and \textbf{30\%} higher NDCG than Google, \ie\ the best performing search engine.   
\end{framed}  
\vspace{-.3cm}

\section{Threats to Validity}\label{sec:threat} 
Threats to \emph{internal validity} relate to experimental errors and biases.
Re-implementation of the existing techniques could pose a threat. However, we used authors' implementation of RACK \cite{saner2016masud} and replicated  \citet{tsc2016} and \citet{cocabu} carefully. We had multiple runs and found their best performances with the authors' adopted parameters which were finally chosen for comparisons. Thus, threats associated with the re-implementation might be mitigated.

Our code corpus (Section \ref{sec:expds}) contains 4,170 documents including 310 ground truth code segments. It is limited compared to a real life corpus (\eg\ GitHub). However, our corpus might be sufficient enough for \emph{comparing} a \emph{generic NL query} with
a \emph{reformulated query} in code retrieval. Please note that our goal is to \emph{reformulate} a query effectively for code search. Besides, we compared with three popular search engines and demonstrated the potential of our query reformulations.
 

Threats to \emph{external validity} relate to generalizability of a technique. Although we experimented with Java based Q \& A threads and tasks, our technique could be adapted easily for other programming languages given that code segments and API classes are extracted correctly from Stack Overflow. 



\vspace{-.1cm}
\section{Related Work\vspace{-.1cm}}\label{sec:related}
\textbf{Relevant API Suggestion:} There have been several studies \cite{portfolio,conngraph,feature,qsynthesis,querythecode,tiqi} that return relevant functions, API classes and methods against natural language queries. \citet{portfolio} employ natural language processing (NLP), PageRank and spreading activation network (SAN) on a large corpus (\eg\ FreeBSD), and identify functions relevant to a given query.
Although they apply advanced approach for function ranking (\eg\ PageRank), their candidate functions were selected using simple textual similarity which is subject to vocabulary mismatch issues \cite{vocaprob}. On the contrary, we apply pseudo-relevance feedback, PageRank and TF-IDF for selecting the candidate API classes.
 \citet{conngraph} apply sophisticated graph mining techniques and return relevant API elements as a connected sub-graph. However, mining a large corpus could be very costly. \citet{feature} mine API documentations and feature history, and suggest relevant methods for an incoming feature request. However, this approach is project-specific and does not overcome the vocabulary mismatch issues. \citet{saner2016masud} apply two heuristics derived from keyword-API co-occurrences in Stack Overflow Q \& A threads, and attempt to counteract the vocabulary mismatch issues during API suggestion. Unfortunately, their approach suffers from low precision due to the adoption of simple co-occurrences. On the contrary, we (1) exploit query-API co-occurrence using a skip-gram based probabilistic model (\ie\ \emph{fastText} \cite{w2vec,fasttext}), and (2) employ pseudo-relevance feedback, Borda count and PageRank algorithm, and thus, (3) provide a novel solution that partially overcomes the limitations of earlier approaches. \citeauthor{saner2016masud} is the most closely related work to ours in API suggestion.  We compare ours with this work, and the detail comparison can be found in Section \ref{sec:eval-api}.
 \citet{qsynthesis} accept free-form NL queries, perform natural language processing, statistical language modelling on source code and suggest relevant method signatures. There exist other works that provide relevant code for natural language queries \cite{lopes1,spotting,sourcerer,nlp2code}, test cases \cite{codegenie,testcase2,s6}, structural contexts \cite{holmes}, dependencies \cite{suade}, and API class types \cite{parseweb,mapo}. On the contrary, we collect relevant API classes for free-form NL queries by mining crowd generated knowledge stored in Stack Overflow questions and answers.
 

\textbf{Query Reformulation for Code Search:} 
Several earlier studies \cite{hillicse09,querythecode,qsynthesis,active,pwordnet,tiqi,li2016,tsc2016,cexchange} reformulate a natural language query to improve the search for relevant code or software artefacts. \citet{hillicse09} expand a natural language query by collecting frequently co-occurring terms in the method and field signatures. Conversely, we apply a different context (\ie\ Q \& A pairs) and a more sophisticated co-occurrence mining (\eg\ skip-gram model). \citet{pwordnet} expand a search query by using part of speech (POS) tagging and WordNet synonyms. \citet{thesaurus} combine WordNet and test cases in the query reformulation. However, WordNet is based on natural language corpora, and existing findings suggest that it might not be effective for synonym suggestion in software contexts \cite{semantictool}. On the contrary, we use a software-specific corpus (\eg\ programming Q \& A site), and more importantly, apply relevant API classes to query reformulation. \citet{active} employ relevance feedback from developers to improve code search. Recently, \citet{tsc2016} collect pseudo-relevance feedback from Stack Overflow, and reformulate a natural language query using Rocchio's method. However, their suggested terms are natural language terms which might not be effective enough for code search given the existing evidence \cite{koderlog}.
Another contemporary work \cite{cocabu} simply relies on Lucene to identify appropriate program elements from Stack Overflow answers for query reformulation.
On the contrary, we employ PRF, PageRank, TF-IDF, Borda count and extra-large data analytics, and provide relevant API classes for query reformulation. 
The above two works are the most closely related to ours. We compare with them empirically, and the detail comparison can be found in Section \ref{sec:eval-qr}.
There exist other studies that search source code  \cite{querythecode,nlsupport}, project repository \cite{li2016}, and artefact repository \cite{tiqi} by reformulating natural language queries. There also exist a number of query reformulation techniques \cite{gayg,refoqus,sisman,kevic,kevicdict,saner2017masud,ase2016masud,observed,shepherd} for concept/feature/bug/concern location. However, they suggest project-specific terms (\eg\ domain terms \cite{domainterm}) rather than relevant API classes (like we do) for query reformulations. Hence, such terms might not be effective enough for code search on a large corpus that contains cross-domain projects.

In short, we meticulously bring together \emph{crowd generated knowledge} \cite{tsc2016}, \emph{extra-large data analytics} \cite{fasttext}, and several IR-based approaches to effectively solve a complex Software Engineering problem, \ie\ query reformulation for code search, which was not done by the earlier studies. Our query reformulation technique can also be employed on top of existing code or web search engines for improved code search (\ie\ RQ$_5$).



\section{Conclusion and Future Work}\label{sec:conclusion}
In this paper, we propose a novel technique--NLP2API--that reformulates a natural language query for code search with relevant API classes. We mine Stack Overflow Q \& A threads,  and employ PageRank, Borda count and extra-large data analytics for identifying the relevant API classes. Experiments with 310 code search queries report that our technique (1) suggests ground truth API classes with 48\% precision and 58\% recall for 82\% of the queries, and (2) improves the given search queries significantly through reformulations. Comparisons with three state-of-the-art techniques and three popular search engines 
not only validate our empirical findings but also demonstrate the superiority of our technique. In future, we plan to investigate the potential of our skip-gram model based on Stack Overflow for project-specific code search (\eg\ concept location, bug localization).

\textbf{Acknowledgement:} This research was supported by Saskatchewan Innovation \& Opportunity Scholarship (2017--2018), and
the Natural Sciences and Engineering Research Council of Canada (NSERC). 

\balance

\bibliographystyle{plainnat}
\setlength{\bibsep}{0pt plus 0.3ex}
\scriptsize
\bibliography{sigproc}  
%
%
\end{document}